\documentclass[twocolumn,pre,aps]{revtex4-1}

\usepackage{amsmath}
\usepackage{amssymb}
\usepackage{color}

\usepackage[pdftex]{graphicx}
\usepackage{subfigure}
\usepackage{color} 

\begin{document}

\title{Improved estimation of anomalous diffusion exponents in single particle tracking experiments}

\begin{abstract}

The Mean Square Displacement is a central tool in the analysis of Single Particle Tracking experiments, shedding light on various biophysical phenomena. Frequently, parameters are extracted by performing time-averages on single particle trajectories followed by ensemble averaging. This procedure however, suffers from two systematic errors when applied to particles that perform anomalous diffusion. The first is significant at short time lags and is induced by measurement errors. The second arises from the natural heterogeneity in biophysical systems. We show how to estimate and correct these two errors and improve the estimation of the anomalous parameters for the whole particle distribution. As a consequence we manage to characterize ensembles of heterogeneous particles even for rather short and noisy measurements where regular time averaged mean square displacement analysis fails. We apply this method to both simulations and in vivo measurements of telomere diffusion in 3T3 mouse embryonic fibroblast cells. The motion of telomeres is found to be subdiffusive with an average exponent constant in time. Individual telomere exponents are normally distributed around the average exponent. The proposed methodology has the potential to improve experimental accuracy while maintaining lower experimental costs and complexity.
\end{abstract}

\author{Eldad Kepten}, 
\email{Eldad.Kepten@biu.ac.il}
\author{Irena Bronshtein}
\author{Yuval Garini}, 
\\
\affiliation{Physics Department $\&$ Institute of Nanotechnology, Bar Ilan University, Ramat Gan, Israel}

\maketitle

\section*{Introduction}

The measurement of single particle tracking (SPT) trajectories is an increasingly popular approach in the study of biophysical systems \cite{BarkaiGariniPhysToday,SaxtonReview}. This measurement can be performed both in vivo or in vitro. Advances in biochemical labelling, nano particle synthesis and the use of fluorescent proteins now allows to label and track almost any biological entity such as, RNA in bacteria \cite{GoldingRNA}, nano-beads in a polymer solution \cite{F-actin}, membrane dynamics \cite{Diego}, and specific nuclear entities \cite{Bronstein,NuclearProteins}. In most biophysical systems, the motion of the tracked particle is stochastic - thus further analysis is necessary before claims regarding the underlying nature of the system can be made. 

Perhaps the most basic analysis technique is the Mean Square Displacement (MSD) in which one calculates the average square of the distance $x(\Delta)$ past by the particle by measurement time lag $\Delta$. This can be performed in two ways. The first is the ensemble averaged MSD, in which the averaging is performed over the displacements of $N_p$ particles, at time $t$ since the beginning of the experiment, $<x_{(t)}^2>=\frac{\sum x^2(t)}{N_p}$. The second is the time averaged MSD (TAMSD) in which all displacements from a single particle trajectory over a time-lag $\Delta=n\tau$ are averaged, where $\tau$ is the time-interval between two consecutive measurements of the particle coordinates and $n$ is an integer. For a trajectory with a total of $N$ measurement time points, the TAMSD is:

\begin{equation}\label{eqn:TAMSD}
\overline{\delta^2_\Delta}=\dfrac{\sum\limits_{m=0}^{N-n}(x[m\tau+\Delta]-x[m\tau])^2}{N-n}
\end{equation}

Stochastic processes can be characterized according to their MSD time dependency. If the MSD depends linearly on time, i.e $<x_{(t)}^2>=2Dt$ or $\overline{\delta^2_\Delta}=2dD\Delta$ (for a $d$-dimensional process), the process is termed normal diffusion, where D is the diffusion constant. Normal diffusion appears, for example, in Brownian motion. Other MSD behaviours are termed anomalous diffusion  \cite{Bouchaud,KlafterMetzler}. 

If the MSD is described by a power law, $<x^2>=D_\alpha t^\alpha$ or $\overline{\delta^2_\Delta}=D_\alpha\Delta^\alpha$ then a further distinction is made. When $1>\alpha>0$ the process is called subdiffusion and for $\alpha>1$ the process is called superdiffusion. $D_\alpha$ is the anomalous diffusion constant and $\alpha$ is the anomalous exponent. The calculated functions TAMSD and EAMSD do not have to show the same behaviour and they may depend on the analysis time  $t$ (or time lag $\Delta$)\cite{SaxtonTrans}.

Anomalous diffusion may result from diverse physical and mathematical origin. The most popular models are the Generalized Langevin equation (GLE) and Fractional Brownian motion (FBM) which can describe the motion of many body systems and in viscoelastic media \cite{LutzGLE,WangTokuyama,Kou,WeberVisco,LizanaGeneral,DengBarkai}; Continuous Time Random Walks (CTRW) which can describe biological trapping \cite{BelBarkai,HeBarkai,SaxtonTrans}; and fractal landscapes which describe, for example, highly obstructed motion \cite{bancaud,szymanski}. Since these models can give rise to similar anomalous exponents, a variety of mathematical tests have been developed to distinguish between them  \cite{MeanMaxExt,Condamin,ScatterMSD,magdziarz,jeonmetz10,BurovPCCP}.Some of these methods have been applied to both experimental and simulated systems \cite{Kepten2011,Diego,WeberVisco,BurneckiKepten2012}.

In order to reach adequate conclusions regarding the nature of the underlying system, it is important to extract accurate parameters that characterize the system. For example, the dynamics of various polymers can be described by GLE that can result from different anomalous models, each with a different anomalous exponent. A phantom Rouse chain will have $\alpha=0.5$ while a Zimm model gives rise to $\alpha=2/3$. Various chain properties can alter this value \cite{LizanaGeneral,granek}, and in the long time domain, all chain models behave similarly according to the crowding of the media\cite{weberPolymer}. Therefore, it is clear that by identifying both the mathematical model and the correct anomalous exponent, one gains deeper understanding of the physical process. 

Identifying the anomalous exponent of a trajectory is not a simple task. Since the process is stochastic, MSDs always have a random nature that depends on the amount of measured time-points. Less time points in the trajectory results in a wider distribution of anomalous exponents that will be found when analyzing the individual particles TAMSDs.  For the time averaged MSD, a reduction in the uncertainty can be achieved by increasing the amount of time-points either by faster acquisition or longer measurements. However, this is usually difficult due to the necessity to use rather long integration-times with the photo-detectors, photo-bleaching that limits the possible acquisition time, induced photo-damage and cell migration. Even if the measurement rate is increased, the MSD will still suffer from erratic behavior at longer time-lags, preventing the extraction of the large time-lag anomalous exponent. Furthermore, previous studies have shown that high rate measurements results in an offset towards lower anomalous exponents due to localization errors \cite{MichaletMeasureNoise,michaletOptimalEst}.  

One possible solution is to measure the trajectories of many particles and perform an ensemble average over the time averaged MSDs, $\left\langle \overline{\delta^2_\Delta}\right\rangle$. This procedure averages out the errors of the individual MSDs and a better estimation at long times is achieved. Biological systems, however, are characterized by heterogeneity of both the tracked particles and the surrounding medium. Therefore, even if a smooth EA-TAMSD is obtained, it cannot give information regarding the individual particles. For some processes like FBM, the non-Markov nature of the increments has prevented the estimation of the individual TAMSD variance around the EA-TAMSD, as was obtained for normal Brownian diffusion \cite{QianSPT}. Finally, as we show later, performing an ensemble average over the TAMSDs does not give the average particle anomalous exponent.  Ultimately, to our knowledge, the success of characterizing the anomalous exponent distribution of the tracked particle population and its time dependency is until now left to the quality of the individual trajectories.

We emphasize that we study the case of heterogeneity between particle's individual anomalous characteristics and not spatial or temporal differences along a specific trajectory (i.e. switching of diffusion modes). The heterogeneity we study where each particle has constant characteristics, can arise from either biophysical or statistical measurement origins. For example the viscoelastic media around the particle may change the anomalous exponent of the stochastic process, and each particle may be in a different viscoelastc domain. Statistical limitations can also give rise to such interparticle heterogeneity. If every particle has the same anomalous exponent of $\alpha$, then limited sampling of their trajectories may lead to an inaccurate estimation of this $\alpha$ and a heterogeneous population would appear. For time correlated processes such as FBM, this statistical spread is expected to be higher due to the correlations in the time averaging scheme. In both of these cases, the distribution of anomalous exponents should be quantified for later study of its origin.

The main purpose of this work is to find the distribution of anomalous exponents of the TAMSDs in a population of tracked particles and this distribution's possible dependence on the time-lag. We assume that the system is limited by localization errors and that particles are heterogeneous (as is commonly found in biophysics).  Our study is limited to ergodic processes and the study of non ergodic and weakly non ergodic processes such as CTRW is left for future study.

Notice that the analysis is performed for one dimensional processes. Usually in two and three dimensional processes, the different directions are assumed to be independent. Thus in order to extend the results of this work to higher dimensions one should only multiply the diffusion coefficient by the dimension, i.e. $MSD_{3D}=3\cdot MSD_{1D}$. Analysis of the existence of such correlations or a fundamental differences between directions is of course an important step in the analysis of any experimental system.

First, we study two systematic errors that are expected to arise in many biophysical SPT measurements. The first (section A) is an offset due to the limited spatial precision (localization accuracy). This issue has been presented in previous works \cite{MartinApparentSub,weberVCF}. We shall expand the analysis of this issue highlighting its importance in the current context and providing a method for its correction.  Section B presents the systematic offset in the ensemble averaged TAMSD caused by the heterogeneity of the individual particle's behaviour. This offset prevents the accurate assessment of the anomalous exponent's time dependency. To the best of our knowledge this subject has not been studied before. In section C we show how to characterise the distribution of the anomalous exponents when there are no localization errors (or they are negligible).

In section D we show how to combine the previous corrections and present a stepwise algorithm for the accurate assessment of the individual anomalous exponents and their time dependency. This algorithm is applied successfully to simulated anomalous diffusion data which includes both localization errors and particle heterogeneity. Finally in section E, we analyse actual SPT experimental data of telomeres \cite{blackburn} diffusion in the nucleus of 3t3 - mouse embryonic fibroblast cells. We show how the naive MSD analysis leads to an erroneous time dependent change of the anomalous exponent, while the new analysis algorithm described here shows that the anomalous exponents are constant, albeit heterogeneous. 

We conclude by emphasizing the potential of the new technique to reduce experimental time-lengths and complexity and to improve the accuracy of the obtained results.

\section*{Results and Discussion}

\subsection{Short time $\alpha$ deviation}

In SPT measurements, the location $x_{(t)}$ can only be measured with a varying error term $N_{(t)}$ that depends on various factors including the point spread function (PSF) of the optical system, the acquisition rate, and other optical and mechanical limitations \cite{MichaletMeasureNoise,michaletOptimalEst,MartinApparentSub,weberVCF}. In general, $N_{(t)}$ is a random variable and its distribution depends on $t$, $x_{(t)}$ and hidden variables such as the hight of the particle above the imaging plain . However, since it impossible to model or measure all these contributions, and since they may vary between experiments, it is customary and beneficial to take a typical stationary probability distribution of this added error, $P(N)$. 

We further assume, based on the central limit theorem,  that $P(N)$ is a normal distribution with zero mean and a standard deviation - $\rho$ . Thus each measurement is actually the sum of the true position and the normally distributed noise: $\hat{x}_{(t)} = x_{(t)} + N_{(t)}$ (throughout this work, 'hats' will represent quantities that include a measurement error). An excellent analysis of the consequences of this additional noise term to the estimation of the anomalous exponent from MSDs is given in \cite{MartinApparentSub}. In this analysis, the authors present the apparent anomalous exponent, here designated $\alpha_N(\Delta)$, and the typical time-lag for which $\alpha_N(\Delta)\approx \alpha$. In this section we go a step further and present a method to correct for this shift in the anomalous exponent even for short measurement time-lags (i.e. when $\alpha_N(\Delta) < \alpha$ ).

Using the fact that the localization error and the particle location are uncorrelated, the apparent MSD for the $i$'th particle is given by \cite{MartinApparentSub} - $\overline{\hat{\delta}^2_\Delta}_i=D_{i,\alpha_i}\Delta^{\alpha_i} + 2\rho^2$. Where we have added the subscript $i$ to designate the possible variability between measured particles and experiments.

In order to extract the anomalous exponent one calculates what is commonly termed the dynamic functional, $\varphi$, by taking the derivative of the logarithm of the MSD according to the logarithm of time, $\varphi= \dfrac{\partial\log\overline{\delta^{2}}}{\partial\log\Delta}$. For sub- and super-diffusion, one expects $\varphi=\alpha$. However, in the case described here, the added noise term will give an erroneous anomalous exponent ,$\varphi=\alpha_N$ \cite{MartinApparentSub}:

\begin{equation}\label{eqn:EffAlpha_N}
\alpha_N(\Delta) = \alpha \dfrac{1}{1+2\rho^2/(D_\alpha\Delta^{\alpha})}
\end{equation}

We find it instructive to present the relative error of the measured dynamic functional, $\epsilon_N = (\alpha_N-\alpha)/\alpha$, in terms of the  quantities, $\overline{\hat{\delta}^2_\Delta}$ and $\rho$. After a short manipulation we find:

\begin{equation}\label{eqn:Error_Noise}
\epsilon_N = -\dfrac{1}{1+(\overline{\hat{\delta}^2_\Delta}-2\rho^2)/2\rho^2}
\end{equation}

From equation ~\ref{eqn:EffAlpha_N} and \ref{eqn:Error_Noise} several facts emerge. First of all, the noise always works to lower the anomalous exponent. Second, the error is time dependent. At very short time intervals, $\Delta\rightarrow 0$, we find $\epsilon_N\rightarrow -1$ and $\alpha_{N(\Delta)}\rightarrow 0$. At long averaging intervals, the error becomes negligible and $\alpha_N\approx\alpha$ (however statistical errors due to low sampling are increased). In addition the relative error is highly dependent on the true $\alpha$ value of the process - which cannot be measured directly. Finally, $\alpha_N$ does not converge to $\alpha$ even for time lags in which the measured (or even the true) MSD are above the noise level. For example if $\overline{\hat{\delta}^2_\Delta}$ is for times higher than $2\rho^2$, $\epsilon_N=0.25$, i.e. a relative error of 25\%.

\begin{figure}[ht]
\begin{center}
\includegraphics[width=4in]{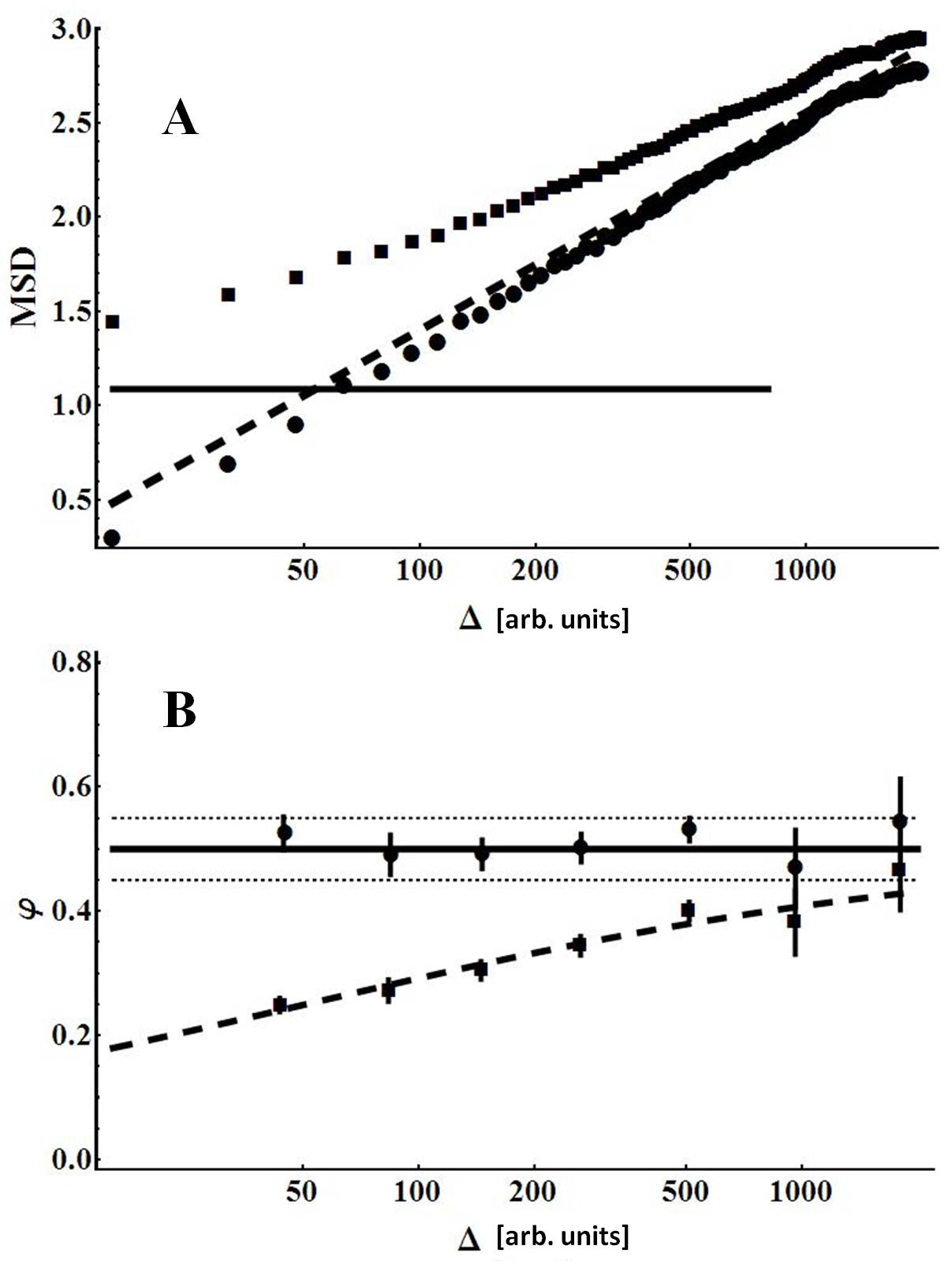}
\end{center}
\caption{
\textbf{Noise causes a systematic error in TAMSDs.} Trajectories of $2^{11}$ time points were simulated with $\alpha=0.5$ together with a localization error of $\rho=1.2$. A) The TAMSD (squares) and  noise corrected MSD (circles) of one of these trajectories are shown together with the theoretical TAMSD without measurement noise (dashed rising line). The noise TAMSD is shown as a parallel continues line. As predicted, the regular TAMSD shows an increasing error at small $\Delta$ while the noise corrected MSD presents a good fit. B)The dynamic functional of the TAMSD (squares) was calculated at various time points and was found to follow the predicted erroneous values (dashed line). The dynamic functional of the noise corrected MSD (circles) on the other hand closely follows the simulated anomalous exponent (continuous line, dotted lines are $\pm 10\%$ of $\alpha$). Since a finite amount of time-points is used for each TAMSD, values of $\varphi$ vary slightly between individual MSDs. The presented points are the average value with standard error bars.
}
\label{fig:NCcorrect}
\end{figure}

Figure \ref{fig:NCcorrect}A shows the TAMSD of a single particle with $\alpha=0.5$ and with noise of magnitude $\rho=1.2$. The deviation from the theoretical prediction is seen clearly at small $\Delta$ values. Notice that as this is a single particle track, it is not smooth. The prediction of equation \ref{eqn:EffAlpha_N} is presented in figure \ref{fig:NCcorrect}B together with the average dynamic functional calculated from ten individual tracks and a clear fit is seen. We stress that averaging a large quantity of tracks will not eliminate this systematic error as it does not originate from limited sampling.

This systematic error can be significantly corrected with simple experimental measures. Suppose a control measurement is performed in order to assess the magnitude of the noise. This is actually a necessary step in any experimental system and can be performed by various means. In live cell measurements, for example, the measurement noise is sometimes classified by measuring a chemically fixated cell \cite{Bronstein}. The static particle data can then be analysed to give the MSD of the localization error alone, $2\rho_c^2$, which should be constant in time and is expected to obey $\rho_c\approx\rho$. Note that this control measurement does not estimate effects such as motion blur. However, these are typically much smaller in SPT experiments, especially with short exposure times  \cite{MichaletMeasureNoise}. One can also estimate $\rho_c$ with the use of simulations \cite{MartinApparentSub}

After finding the magnitude of the noise, one can calculate the noise corrected MSD,
\begin{equation} \label{NCMSD}
\nu_{(\Delta)}=\langle\overline{\hat{\delta}^2_\Delta}\rangle-2\rho_c^2
\end{equation}
The triangular brackets are an ensemble averaging on the various trajectories. This ensemble averaging is crucial due to the the random nature of both the stochastic process and the localization error magnitude (which varies between experiments). The variability may cause the subtraction to result in negative values for individual particle MSDs. Such negative values will prevent the calculation of the dynamic functional since the logarithm of a negative value is impossible in this context. Negative values are quickly eliminated with ensemble averaging since $\langle\overline{\hat{\delta}^2_\Delta}\rangle = \langle \overline{\delta^2_\Delta}\rangle + \langle\rho^2\rangle$.
Hence one finds $\nu_{(\Delta)}\approx \langle\overline{\hat{\delta}^2_\Delta}\rangle$ and for an an ensemble of particles with identical $\alpha$ values, $\nu_{(\Delta)}\approx\langle D_\alpha\Delta^\alpha\rangle$. Now, calculation of the dynamic functional will give $\alpha$ without the effect of the localization error.  

This correction is applied to the previous simulations and presented in figure \ref{fig:NCcorrect}A and B. It is clear that the noise corrected MSD accurately retrieves the anomalous exponent of the simulated process even when the noise is of the same magnitude as the process MSD. This experimental scenario was previously impossible to study without large errors. In theory, the noise corrected MSD enables the measurement and analysis of anomalous diffusion down to unlimited short time lags. 

Notice that if $\alpha$ is varying between particles, this technique will still give $\langle \overline{\delta^2_\Delta}\rangle$ even though the relative error, $\epsilon_N$ can vary significantly between particles at each time point.

\subsection{Long time $\alpha$ deviation}

\begin{figure}[!ht]
\begin{center}
\includegraphics[width=6in]{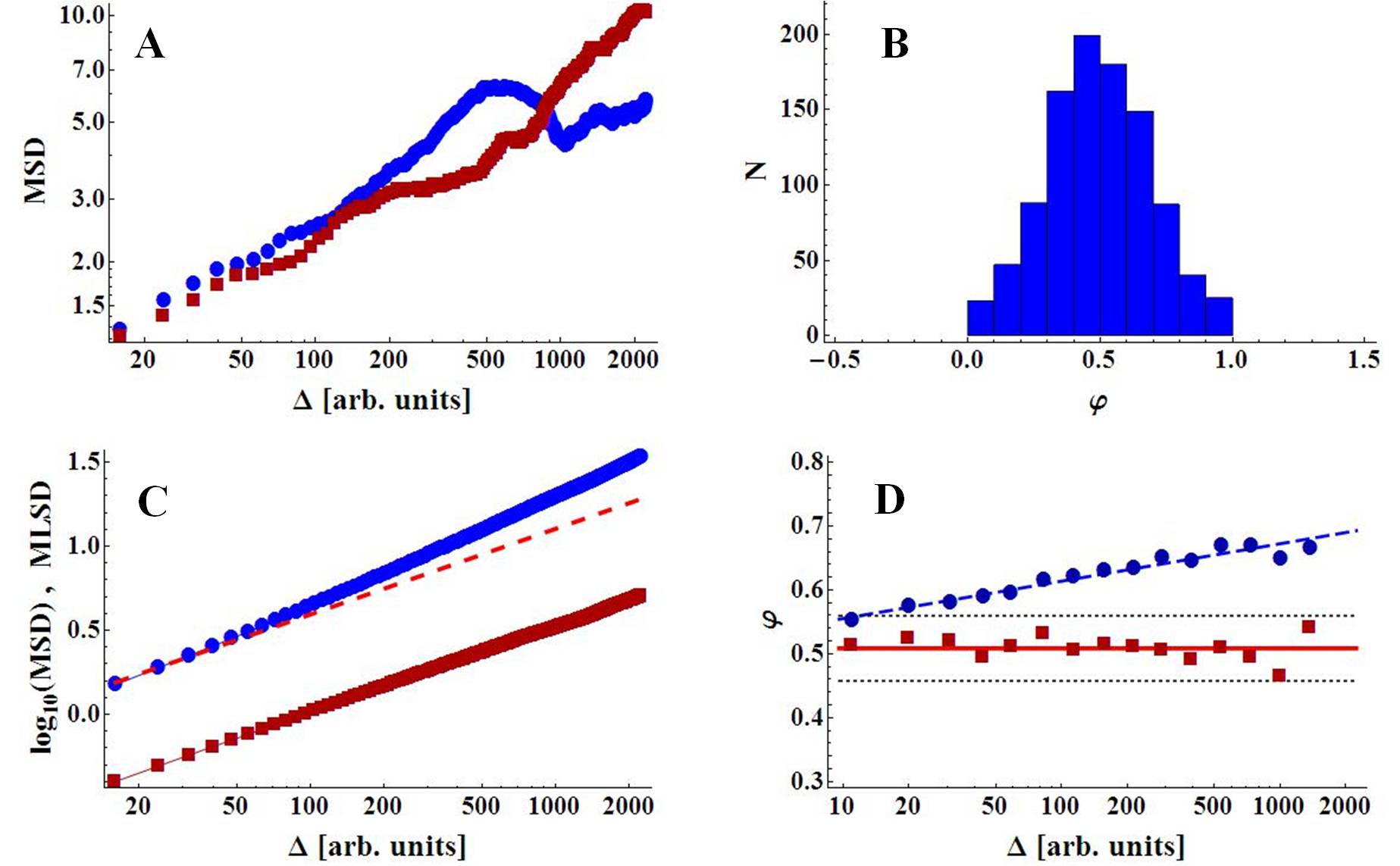}
\end{center}
\caption{
\textbf{(color online) Correction of the ensemble variation error.} For $2^{10}$ particles, an anomalous exponent was taken from a normal distribution with $\mu_\alpha=0.5$ and $\sigma_\alpha=0.2$. $2^{10}$ time points were taken for each trajectory. A)The TAMSDs of two representative particles are shown. The roughness of the path and random deviations from a straight line can be seen. B)The histogram of $\alpha$ values simulated. C) The regular EA-TAMSD (circles, blue), the MLSD (squares, red) and the theoretical prediction for the average particle (dashed red line). The deviating slope of the MSD can be seen while the MLSD is parallel to the theoretical average particle (compare slope of red squares and blue circles to dashed line). D) The dynamic functional was calculated for the EA-TAMSD (circles, blue) and MLSD (squares, red) at constant intervals on the logarithmic scale. The EA-TAMSD dynamic functional is linearly increasing (dashed blue line), while the MLSD is around the simulated $\mu_\alpha=0.5$ (red line). Dotted black lines denote 10\% errors from $\mu_\alpha$.
}
\label{fig:SpreadGraph}
\end{figure}

When calculating the TAMSD of a single particle, for large $\Delta$ values or short tracks, there are not many time points to average over. Hence the TAMSD suffers from high variation and random errors \cite{QianSPT}. These errors are obvious for example in Figure \ref{fig:SpreadGraph}A where it is impossible to analyse the TAMSD after the first few time points.  

A possible way to mitigate this problem in the case of normal diffusion is to take the ensemble average of many TAMSDs (EA-TAMSD) of different tracks, $\left\langle\overline{\delta^2_\Delta}\right\rangle$. This method is adequate for normal diffusion or for identical particles, retrieving accurate average diffusion parameters for the system.

It fails, however, for the general case of anomalous diffusion where the tracers or their individual vicinity are not identical. In other words, the ensemble average TAMSD of a heterogeneous population does not represent the behaviour of the typical particle. As we show below, the dynamic functional extracted from $\left\langle\overline{\delta^2_\Delta}\right\rangle$ is different from the average anomalous exponent and shows a time dependency. 

\begin{figure}[!ht]
\begin{center}
\includegraphics[width=4in]{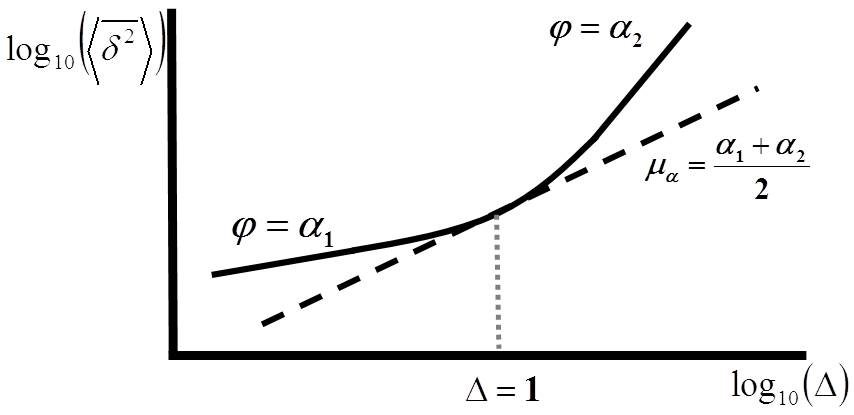}
\end{center}
\caption{
\textbf{An illustration of a heterogeneity induced time dependent anomalous exponent.} For two particles with the same diffusion coefficient and $\alpha_2>\alpha_1$, the ensemble averaged TAMSD shows a time dependent anomalous exponent (Black line). Notice that this is a log-log scale, and so the dynamic functional $\varphi$ (equal to the instantaneous anomalous exponent) is the slope of the graph. The dotted line represents the expected MSD according to the average exponent, $\dfrac{\alpha_1+\alpha_2}{2}$. Naively studying the ensemble averaged MSD would lead to the conclusion of a time dependent anomalous exponent. 
}
\label{fig:Example}
\end{figure}

We start by demonstrating the problem with a simple case of two particles only. Assume that both of them have a diffusion constant of $D_\alpha^{1,2}=1$ and that $\alpha_2>\alpha_1$. It is clear that the average anomalous exponent is $ \frac{\alpha_1+\alpha_2}{2}$. However, the average TAMSD is $ \frac{\Delta^{\alpha_1}+\Delta^{\alpha_2}}{2}$(omitting the $D_\alpha=1$). Upon extracting the dynamic functional from this average TAMSD, one will find that at large time lags, i.e. $\Delta>>1$, the average MSD behaves according to $\varphi\approx\alpha_2 $ while at $\Delta<<1$ the behaviour is governed by $\varphi\approx\alpha_1$ (figure \ref{fig:Example}). 

In the general case, we assume a large ensemble of particles with a normal distribution for $\alpha$, with mean $\mu_\alpha$ and standard deviation $\sigma_\alpha$. Since $\alpha_i$ must be positive, such a distribution must be limited at zero, i.e. $\alpha_i>0$. However, this limitation is negligible unless $\sigma_\alpha\simeq\mu_\alpha$ and can be quantified, see appendix A. We also take $D_{\alpha,i}\sim P(D)$ with an average $\mu_D$. However, $D_{\alpha,i}$ and $\alpha_i$ are not correlated. As we will see later, the diffusion constants do not affect the accurate extraction of the anomalous exponent. This stems from our interest in the power law behaviour at long times, dimming the diffusion constants influence negligible. We also assume no measurement error at this stage of our analysis. 

One could come up with distributions of parameters for which the approximation to a Gaussian fails at certain conditions or perhaps even $\mu_\alpha$ and $\mu_D$ do not exist. We leave these distributions to future work and emphasize that they are not the common experimental scenario. 

For the above distribution of anomalous exponents, with each trajectory behaving according to the theoretical prediction $\overline{\delta^2_\Delta} = D_{\alpha,i}\Delta^{\alpha_i} $, then the distribution of TAMSD is almost Log-Normal (see Appendix A). Taking an integral over the distribution through a series of variable changes we find that the ensemble averaged TAMSD behaves according to:

\begin{equation} \label{eqn:NDmean}
   \left\langle \overline{\delta^2_\Delta}\right\rangle = \mu_D\Delta^{\mu_\alpha}  \cdot \exp(\dfrac{1}{2}\sigma_\alpha^2 \ln^2(\Delta))
\end{equation} 

It is clear that there is a time dependent offset from the average anomalous exponent $\mu_\alpha$. In order to find the time dependent anomalous exponent,$\alpha_S(\Delta)$, we again calculate the dynamic functional ($\varphi=\alpha_S$). One finds an effective average anomalous exponent of: 

\begin{equation} \label{eqn:NDslope}
   \alpha_S = \mu_\alpha +\sigma_\alpha^2 \ln(\Delta) 
\end{equation} 

From equation ~\ref{eqn:NDslope} we find the difference in the dynamic functional between two time lags. For $\Delta$ and $\Delta_0$, and defining $d_\alpha = \alpha_{(\Delta)}-\alpha_{(\Delta_0)}$ we find:

\begin{equation} \label{eqn:EnsembleError}
d_\alpha = \sigma_\alpha^2 \ln(\frac{\Delta}{\Delta_0})
\end{equation}

Notice that $d_\alpha$ is independent of the diffusion constant and shows a logarithmic dependency on $\Delta$.  

To test this result, 1000 particles with a distribution of individual anomalous exponents according to $\mu_\alpha=0.5$ and $\sigma_\alpha^2=0.2$ were simulated for 1000 time points. The retrieved distribution of anomalous exponents is shown in Figure \ref{fig:SpreadGraph}B. Two graphs of the MSDs of individual particles are shown in Figure \ref{fig:SpreadGraph}A. The EA-TAMSD of the ensemble is shown in Figure \ref{fig:SpreadGraph}C. Notice that the time lag and MSD are plotted on a logarithmic scale and a sub-diffusive MSD is shown as a straight line with slope $\alpha$. Evidently, the EA-TAMSD has a varying slope, as expected from equation \ref{eqn:NDslope}. The dynamic functional extracted at each time-lag is seen in Figure \ref{fig:SpreadGraph}D.  It is clear that the error in the dynamic functional grows linearly with the logarithm of the time lags. Notice that these simulations were preformed without the addition of measurement noise, and thus the single source of systematic error is the distribution of anomalous exponents.

\subsection{Extracting $\mu_{\alpha}$ and $\sigma_{\alpha}$ for data without localization error}

The time dependency of the dynamic functional arises from the attempt to compare between different powers of time. Hence, one should try to directly average the exponent values while cancelling the exponential time dependency of each trajectory. This can be done by averaging between the logarithm of the exponential quantities - creating a linear dependency in time. For this purpose we present the logarithmic square displacement (with $\Delta=n\tau$ as before):
\begin{equation}\label{eqn:MLSD}
\xi_{(\Delta)}=\log(\dfrac{\sum\limits_{m=0}^{N-n}(x[(m\tau+\Delta]-x[m\tau])^2}{N-n})
\end{equation}

For a process that shows anomalous TAMSDs, this gives for each particle $i$, $\xi_{(\Delta)}=\log(D_{\alpha,i})+\alpha_i\log(\Delta)$. Now, when taking the ensemble average, i.e. the Mean Logarithmic Square Displacement (MLSD), the true average exponent is found - 
\begin{equation}\label{eqn:MLSDcalc}
\langle\xi_{(\Delta)}\rangle=\langle \log(D_{i})\rangle + \mu_\alpha\log(\Delta)
\end{equation}
And specifically, $\varphi=\mu_\alpha$ for all times. The results of the MLSD for the same ensemble of particles in Figure \ref{fig:SpreadGraph}B is also presented in Figure \ref{fig:SpreadGraph}C and D. The accurate reconstruction of $\mu_\alpha$ can be seen.

The MLSD gives more details about the anomalous behaviour of the diffusing particles than just $\mu_\alpha$. One can also extract the standard deviation of the individual anomalous exponents, $\sigma_\alpha$, and thus characterize the whole population.

At each time point, $\Delta$ both the MLSD slope $\mu_\alpha(\Delta)$ and the ensemble averaged TAMSD slope $\alpha_S(\Delta)$ are calculated and the set of errors is extracted as a function of $\Delta$:
\begin{equation} \label{eqn:relative_errors}
\epsilon_i=\alpha(\Delta)-\alpha_S(\Delta)
\end{equation}
This set can then be fitted against $\ln(\Delta)$ according to a small manipulation of equation \ref{eqn:EnsembleError},
\begin{equation}\label{eqn:Fitting_errors}
\epsilon_i=\sigma_\alpha^2\ln(\Delta)-\sigma_\alpha^2\ln(\Delta_0)
\end{equation} 
The slope of this fit is the variation of the anomalous exponents, $\sigma_\alpha^2$. 

Two central issues should be emphasized. First, note that both the MLSD and the ensemble averaged TAMSD are similarly affected by the noise induced error, $\epsilon_N$. Thus, the $\Delta_i$'s used for the fit can extend to small $\Delta$ values, even though $\epsilon_N$ might be significant. This greatly increases the amount of points that can be used for the fit.

Second, in some experimental scenarios, this technique can be much more accurate than fitting each individual TAMSD to a power law. For example, if only a small number of time points are measured, the individual TAMSDs are very erratic. Thus the ability to extract individual anomalous exponents from each track is very low. 

\begin{figure}[!ht]
\begin{center}
\includegraphics[width=4in]{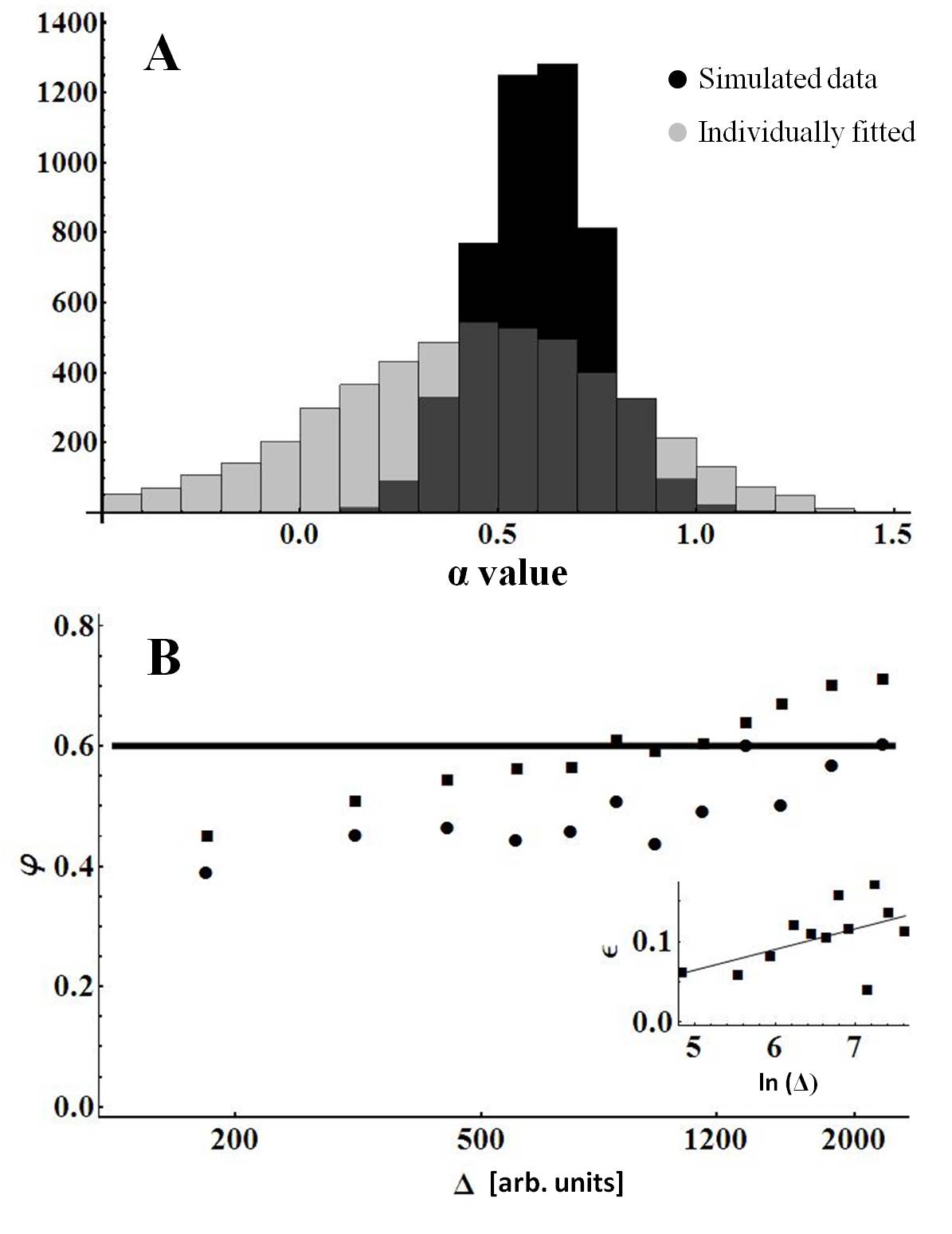}
\end{center}
\caption{
\textbf{Finding the standard deviation of anomalous exponents.} 5000 short trajectories ($64$ time points), were simulated with $\mu_\alpha=0.6$ and $\sigma_\alpha=0.15$. In addition an experimental noise of $\rho=1.5$ was added. A) Trying to find the anomalous exponent for each track individually fails, as a much wider distribution (gray) is found - $\sigma^{ind}_\alpha=0.37$, an error of 150 \%  - with an offset towards lower values, compared to the simulated values (black). B) For 12 time points, the dynamic functional of the EA-TAMSD (square) and MLSD (circle) is calculated. Even though they are both under the simulated average (straight line), the differences can be fitted to a straight line on a logarithmic scale (inset). The resulting standard deviation found was $\sigma_\alpha=0.16$, an error of 7\%.
}
\label{fig:Sigma}
\end{figure}

To test this technique, 5000 particles with $\alpha_i$ taken from a normal distribution with mean $\mu_\alpha=0.6$ and standard deviation $\sigma_\alpha=0.15$ were simulated. To further complicate the analysis, a measurement noise of $\rho=1.5$ was incorporated into the process. For each track, only $64$ time points were taken, thus the ability to fit each anomalous exponent is very low. Figure \ref{fig:Sigma}A shows the histogram of the simulated $\alpha_i$ values and the fitted ones. Two errors are evident. The first is the very wide spread of the fitted distribution, and the impossible negative values found in some cases. Actually, the standard deviation of the fitted $\alpha_i$'s was $\sigma_\alpha^{ind}=0.37$, an error of 150\%.  The second error is the shift of the peak value due to the measurement noise to an average value of 0.4.

Figure \ref{fig:Sigma}B presents the extraction of $\sigma_\alpha$ with the use of the MSD and MLSD. The measurement noise causes the dynamic functional found for the MLSD to be well below the average value of $\mu_\alpha=0.6$. This however does not affect the result, as the standard deviation of alpha values was found to be $\sigma_\alpha^{new}=0.16$, an error of only 7\%. We see that the MSD error estimation technique extracts $\sigma_\alpha$ even for short, noisy trajectories, a task not possible previously. In the next section we show how to correct for the localization error and find $\mu_\alpha$ in addition to $\sigma_\alpha$.

\subsection{General methodology for noisy and heterogeneous trajectories}

We now combine the previous conclusions into a general methodology for the analysis of an ensemble of anomalously diffusing, heterogeneous particles. These particles are measured with an inherent localization error and the goal is to characterise the distribution of anomalous exponents (i.e. $\mu_\alpha$ and $\sigma_\alpha$). The procedure includes the following steps:
\begin{enumerate}
\item Measure the trajectories of all particles and calculate for each $i$'th particle its TAMSD, $\overline{\delta^2_{\Delta,i}}$.
\item Using a control measurement or simulation , extract the standard deviation of the localization error, $\rho_c$.
\item For a set of time-lags $\{\Delta_j\}$ (we find that seven equally spaced time lags are sufficient for a thousand time point trajectory) calculate the local dynamic functional of the ensemble averaged TAMSD, $\alpha_S(\Delta_j)$ (equation \ref{eqn:NDslope}), and the slope of the MLSD from equation \ref{eqn:MLSDcalc} (notice that due to measurement noise this is not $\mu_\alpha(\Delta)$ but rather includes an error according to equation \ref{eqn:Error_Noise}).
\item Using equations \ref{eqn:relative_errors} and \ref{eqn:Fitting_errors} find the standard deviation of anomalous exponents, $\sigma_\alpha$ and the parameter $\Delta_0$. 
\item Calculate the noise corrected MSD $\nu_{(\Delta)}$, equation \ref{NCMSD}. 
\item The final corrected average particle MSD is found by dividing $\nu_{(\Delta)}$ by $\Delta^{d_\alpha}$ for which we have all parameters , see equation \ref{eqn:EnsembleError}: 
\begin{equation}\label{eqn: FinalMSD}
\nu_c(\Delta)=\frac{\nu(\Delta)}{\Delta^{d_\alpha}}
\end{equation}
\item Finally one calculates the slope of $\nu_c(\Delta)$ and finds $\mu_\alpha(\Delta)$ with out the error of the measurement noise or ensemble heterogeneity. 
\end{enumerate}

An assumption is made that $\sigma_\alpha$ is constant with time-lag. If for example there is an increase of $\sigma_\alpha$ then one would find a non linear dependency in equation \ref{eqn:Fitting_errors} on $\ln{\Delta}$. In such case a deeper analysis of the time-lag dependency will be needed. 

This procedure was implemented for a set of 1000 particles, measured at 1024 time points, with $\mu_\alpha=0.6$ and $\sigma_\alpha=0.2$. In addition, a measurement noise with standard deviation $\rho=1.5$ was added. Here too, the analysis of individual trajectories is bound to suffer from random errors. 

\begin{figure}[!ht]
\begin{center}
\includegraphics[width=4in]{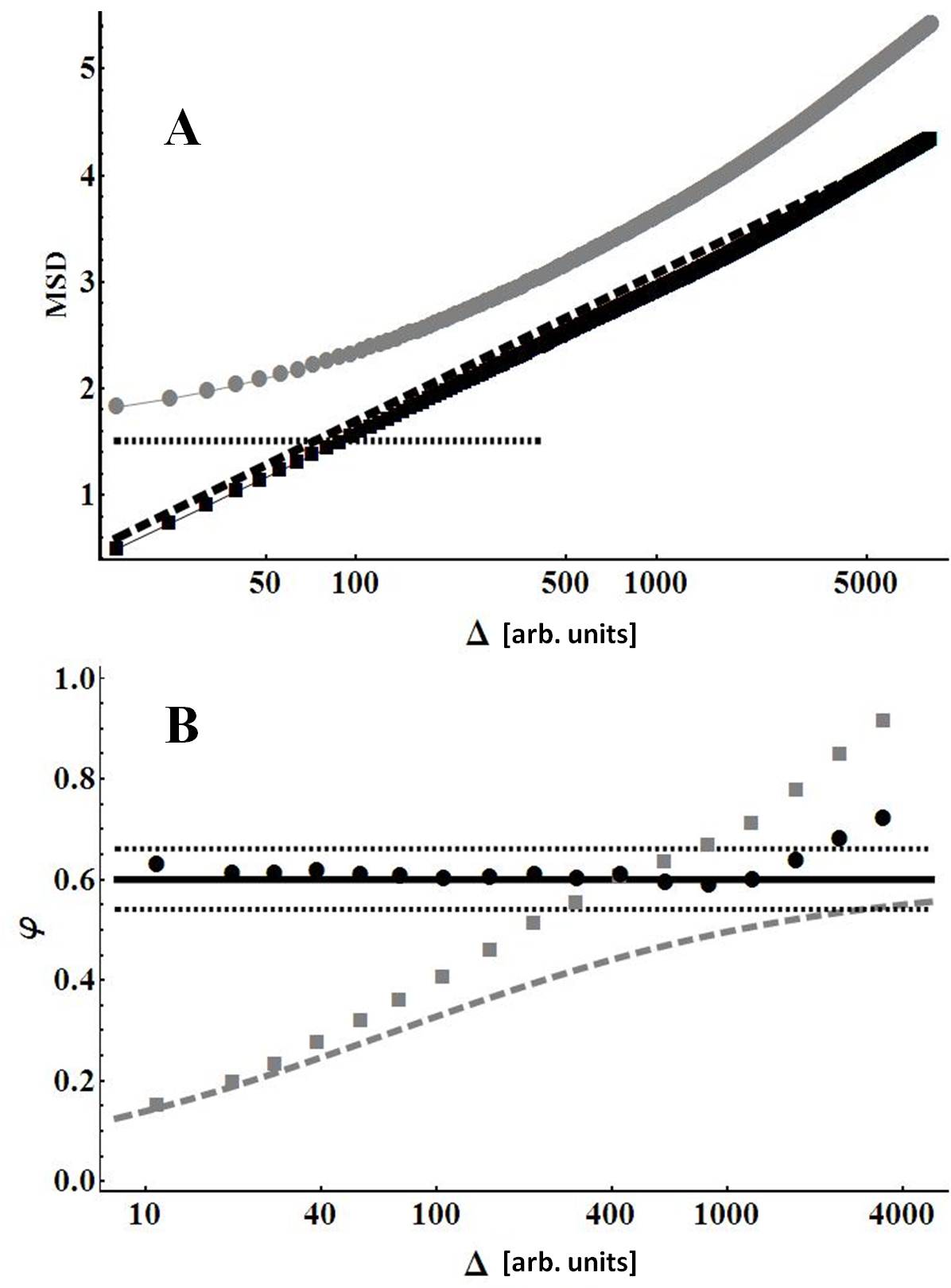}
\end{center}
\caption{
\textbf{Constructing the average particle MSD.} 1000 trajectories of $2^{10}$ time points were simulated with $\mu_\alpha=0.6$, $\sigma_\alpha=0.2$ and an additional noise of $\rho=1.5$ (dotted line). A) The EA-TAMSD (gray) shows strong deviations both in short and long times. The corrected average MSD ($\nu_c$, black squares) is almost identical to the theoretical average particle (dashed line). B) Calculating the dynamic functional for the EA-TAMSD (gray) and $\nu_c$ (black) shows the dramatic improvement. Black continuous line is the ensemble average, $\mu_\alpha=0.6$, dashed gray line is the noise induced error and black dotted lines are 10\% errors from $\mu_\alpha$
}
\label{fig:BothTechniques}
\end{figure}

First, the ensemble distribution error, $d_\alpha$ was extracted by comparing the MSD to the MLSD as stated above. Then,$\nu(\Delta)$, was calculated for each time point. Finally, equation \ref{eqn: FinalMSD} was used to find $\nu_c(\Delta)$. The uncorrected  ensemble averaged TAMSD, $\left\langle\overline{\hat{\delta}^2_\Delta}\right\rangle$ and $\nu_c(\Delta)$ are both presented in Figure \ref{fig:BothTechniques}A with the theoretical TAMSD of the average particle. The significant improvement is clear. Finally, Figure \ref{fig:BothTechniques}B shows the dynamic functional at each time point of both the uncorrected MSD and the final noise and variation corrected MSD, again a significant improvement is seen.

\subsection{Experimental application example}

In order to test our proposed analysis method on actual experimental data, we set to characterize the diffusion of telomeres (the ends of the chromosomes, \cite{blackburn}) in the nucleus of 3T3 mouse embryonic fibroblast cells. The measurements were performed by standard confocal microscopy, according to the same protocol previously published in \cite{Bronstein}. In short, cells are transiently transfected with a plasmid coding for a merged TRF1-GFP protein. These proteins are integrated into the sheltrin complex surrounding each telomere, thus enabling the tracking of approximately a dozen telomeres per cell. We measure two time regimes - a fast regime spanning $100$ seconds at $2Hz$ (200 time points) and a slow regime up to 30 minutes with 100 time points. The motion of telomeres has been previously identified as obeying fractional Brownian motion \cite{Kepten2011,BurneckiKepten2012}.

\begin{figure}[!ht]
\begin{center}
\includegraphics[width=6in]{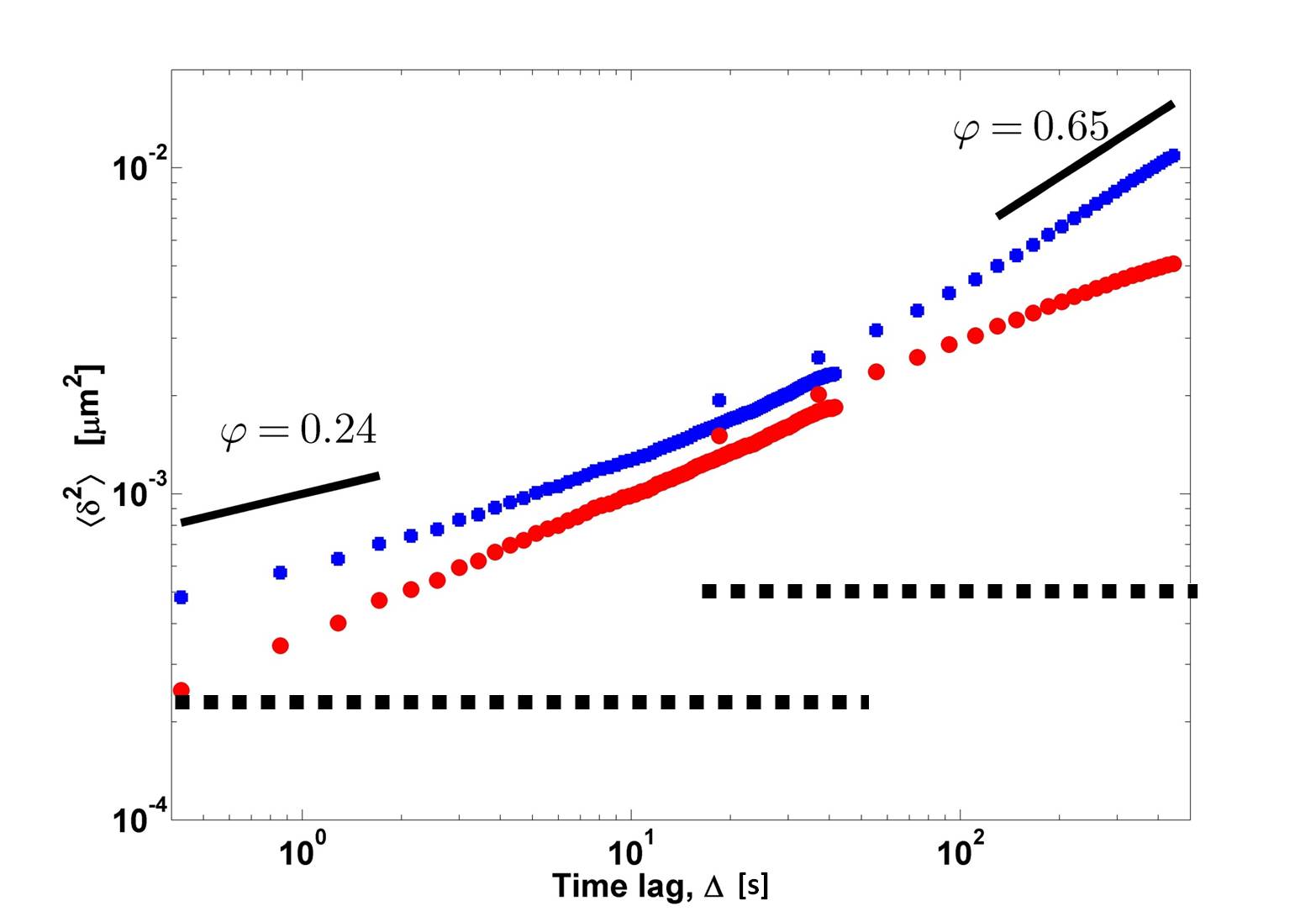}
\end{center}
\caption{
\textbf{(color online) Analysis of telomere diffusion in 3T3 cells.} The ensemble averaged TAMSD (blue squares) showed a time dependent dynamic functional (slope of the graph on a log-log scale). The shortest times displayed $\varphi=0.24$ while the longest times $\varphi=0.65$. Notice that the noise level (dotted black lines) is significantly lower than the MSD. Applying the full analysis algorithm (red circles), i.e. correcting for both the measurement noise and ensemble heterogeneity gave a constant average anomalous exponent of $\mu_\alpha=0.4\pm 0.04$ and an ensemble standard deviation of $\sigma_\alpha=0.1$ and $\sigma_\alpha=0.2$ for the short and long time measurements respectively. This spread may result from biophysical or statistical reasons.
}
\label{fig:Experimental}
\end{figure}

As with the individual simulated trajectories in figure \ref{fig:SpreadGraph}A no information could be extracted after the first few time-lags and any result for the anomalous exponent is highly dependent on the specific time lags taken for the fitting. An ensemble averaging of the TAMSD was performed, and is presented in figure \ref{fig:Experimental}. Such a naive analysis would conclude that the typical anomalous exponent is time dependent (from $\alpha=0.24$ to $\alpha=0.65$) and no information regarding the individual telomeres is obtained. 

Following the steps of our proposed algorithm we first quantify the measurement noise in our system and find it to be $2\rho_c=2.3\cdot 10^{-4} \mu m^2$ for the short time regime and $2\rho_c=5\cdot 10^{-4} \mu m^2$ for the long time regime. We then calculate the differences between the regular EA-TAMSD and the MLSD. We find that for the short and long time regimes, $\sigma=0.1, \Delta_{0}=0.8$ seconds and $\sigma=0.2, \Delta_{0}=27$ seconds, respectively. 

According to the last stage of our methodology, for each time regime we subtract the noise term from the regular EA-TAMSD and divide by the correction term from equation \ref{eqn:EnsembleError}. The final corrected average particle MSD is also presented in figure \ref{fig:Experimental}. The resulting anomalous exponent is almost constant in time, $\mu_{\alpha}=0.4\pm 0.04$ (standard deviation) and the two time regimes are continuous. No significant time dependency is found and the heterogeneity of the individual MSDs is quantified.

We suggest that the heterogeneity in the anomalous exponents originates in a combination of a true biophysical distribution and a statistical randomness due to the short trajectories. A comparison with an ensemble of simulated FBM processes with $\alpha=0.4$ may show the relative roles of these two origins.

It should be noted that the original MSD and the individual single particle MSDs could not give accurate information regarding the distributions of the anomalous exponents at the short time regime even though they are above the experimental noise. This originates from equation \ref{eqn:Error_Noise} which shows that the noise induced error propagates into large time lags. This highlights the importance of always correcting for experimental noise. In addition, it was impossible to assess all longer time-lags of the MSDs due to their erratic behaviour. Evidently, our proposed technique manages to deal with all these effects.

\section*{Conclusions}

In this work we have studied two types of systematic errors in the estimation of anomalous diffusion MSDs, that may hinder attempts to accurately characterise such diffusive systems. These errors, which are the common situation for biophysics, cannot be corrected by increasing the amount of averaged particle locations (either by longer measurements, faster acquisition rates or more particle trajectories). Rather, they are inherent to noisy, heterogeneous systems. Such systematic errors can lead to erroneous conclusions about the physical and biological nature of the system. 

We have devised an experimentally applicable stepwise methodology to overcome the effects of measurement noise and ensemble heterogeneity and extract both the average anomalous exponent and the standard deviation around it. Our methodology was developed for sub and super diffusive ergodic processes such as FBM or obstructed diffusion, under the assumptions of normally distributed measurement error, and a normal distribution of anomalous exponents among tracked particles. Of course, for parameter distributions far from a Gaussian shape this technique will show deteriorated accuracy.

Every experimental measurement has a finite accuracy and incorporates some degree of uncertainty. In SPT experiments, the location of the tracked particle is known up to a random noise term. As was shown before \cite{MartinApparentSub}, this added noise causes a systematic decrease of the measured anomalous exponent, even when the true MSD is much larger than the noise. We have shown how to quantify and correct this error with simple control experiments for any time-lag, even when the error is significant. 

Heterogeneity in biology may stem from a variety of reasons. First and foremost, individual cells vary from each other and even inside each cell, each tracked particle may experience different surroundings, especially in finite time measurements. In addition, biological tracers may differ in their properties and a central goal of the study is to characterize the distribution of parameters in the system. This is also true regarding the anomalous exponents of the individual particles. However, when analysing short tracks, it is difficult to accurately identify the parameters of single particles individually, since the small amount of time-points introduces large errors. Hence, one aspires to reach conclusions from studying ensemble averaged quantities such as the ensemble averaged TAMSD.

We have shown that the distribution of anomalous exponents creates a time-dependent systematic error in the extraction of the average anomalous exponent from the ensemble averaged TAMSD. By using the newly defined MLSD, this error can be corrected. 

With the methodology to combine the noise and the ensemble heterogeneity corrections an accurate MSD of the average particle behaviour is achieved (i.e. the average diffusion coefficient and anomalous exponent). This corrected average MSD suffers less random errors due to the large amount of points averaged upon and does not suffer from the systematic errors of the original ensemble averaged TAMSD. Since this technique gives both the average anomalous exponent and its variance, the whole distribution of anomalous exponents is characterized in a much more accurate way than possible before.

We have also implemented this methodology on experimental results of telomere subdiffusion in 3T3 cells. This system was impossible to accurately analyse by individually fitting particle TAMSDs. Also, a naive analysis using only the ensemble averaged TAMSD would have led to erroneous results of a time dependent anomalous exponent due to measurement noise and heterogeneity of the population. By applying our methodology we where able to characterise the whole distribution of anomalous exponents. Telomeres where shown to have a constant anomalous exponent distributed around the average value of $\mu_\alpha=0.4$. 

The ability to characterise the whole distribution from ensemble averaged quantities was shown to work even for extremely short trajectories, where single particle analysis failed completely. Another option is to measure more data points, however this is usually a complicated experimental challenge which demands more complex and expensive equipment. Our proposed technique enables to improve the characterisation of the single particle distribution by measuring more particles and not longer trajectories. Thus it carries high potential to reduce experimental costs and complexity. 

Quantification of the distribution of anomalous exponents,$\sigma_\alpha$, can shed light on the biophysical function of the system. Since the ensemble includes particles with high anomalous exponents in addition to particles with low exponents, the ensemble covers a larger area than would be expected if all particles behaved according to the average particle. At the same time, particles with low anomalous exponents tend to stay more localized around the origin of the motion. Thus the first group acts as searchers of distant targets while the second group raises the efficiency of interaction near the origin. We therefore see that a distribution of anomalous exponents can be beneficial in various sceneries of biological signalling and interactions.  

The growing number of studies in live cells, new tracer families, faster acquisition rates and larger quantities of tracers per experiment will enable the development of novel biophysical theories. The correct extraction of parameters from the measured data is essential for these theories to be accurate and act as useful predictors of biological processes. Using the tools described herein will enable to better characterize SPT results which in turn will no doubt improve our understanding of biophysical systems, and help to formulate better theoretical models for their description.

\section*{Materials and Methods}

\subsection*{Cell growth and measurement}
Cell growth, preparation and measurement were performed with the same equipment and guide lines as was previously published in \cite{Bronstein}, with a few modifications due to the use of 3T3 mouse embryonic fibroblast cells instead of human U2OS osterosarcoma cells. Specifically, cells where grown in a high glucose medium (4.5 \%). Transfection with the TRF1-GFP plasmid was performed through a standard electrophoresis protocol (BioRad, $160V,\space 1200\mu F,\space\inf\Omega,\space 4mm$ cuvette). Cells were imaged using an inverted Olympus IX-81 fluorescence microscope (UPLSAPO Objective lens 60x,
NA=1.35) coupled to a FV-1000 confocal setup (Olympus). For fast measurements ($2 Hz$, 200 time points) only a single focal plane was imaged. At times longer than $\sim 1$ minute the cells begin to migrate and rotate. Thus a full three dimensional scan was performed in order to correct the telomeres positions for this motion. Image analysis was done with the Imaris image analysis software package.

\subsection*{Simulation of FBM}
Particle tracks where simulated using the MATLAB \textit{wfbm} function. Each path $i$ was built for N time points with a random Hurst index, $H_i$, taken from a distribution with mean $\mu_\alpha$ and variance $\sigma^2_\alpha$ (for FBM, $2H=\alpha$). The paths were then down-sampled by a factor of 16 to give $T$ time points. This down-sampling removes high frequency errors in the data. \textit{Wfbm} uses the algorithm developed by Sellan and Meyer \cite{abryWFBM}, and thus the simulated diffusion constant of each particle obeys $D_{\alpha,i}=\Gamma(1-\alpha)\frac{Cos(\pi\alpha/2)}{\pi\alpha/2}$. In order to simulate paths with a minimal correlation between $\alpha_i$ and $D_{\alpha,i}$ each path was multiplied by $\sqrt{\Gamma(1-\alpha)Sin(\pi\alpha/2)}$. Since $\Gamma(1-\alpha)\Gamma(1-\alpha)\frac{Sin(\pi\alpha/2)Cos(\pi\alpha/2)}{\pi\alpha/2}=1$ this significantly reduces the correlation.
Measurement noise was simulated by adding a random noise increment $N_{i,(t)}$ to each track at each time point. This increment was taken from a stationary normal distribution with a mean of zero and a variance of $\rho^2$ according to the requested noise scenario. 

\subsection*{Path analysis}
Since the purpose of this article is to present new methods for the analysis of anomalous processes, the analysis of simulated data was performed as if no prior data was available. Specifically, a separate Mathematica program was written, which analysed the ``experimental'' paths only with the help of another set of simulated ``experimental noise'' paths with the same $\rho^2$ variance and no FBM process.

\section*{Acknowledgments}
EK would like to thank Eli Barkai for many helpful discussions and directions. This work was supported in part by the Israel Science Foundation grant No. 51/12.

\section*{Appendix A - Calculation TAMSD distribution and $\alpha>0$ influence}
Assuming a normal distribution of $\alpha$, with mean $\mu$ and standard deviation $\sigma$ than, $z=e^\alpha$ has a log normal distribution. Thus $P(z)=(\sqrt{2\pi} \sigma z)^{-1}\cdot \exp(-(\ln(z)-\mu)^2/2\sigma^2)$. We are interested in the distribuion of the TAMSD $y\equiv\hat{\delta^2}=\Delta^\alpha$. Using $y=\Delta^{\ln(z)}$ and $\frac{e^{\log_\Delta(y)}}{\ln(t)}dy=dz$ we find:
\begin{equation*}
P(y)dy=\dfrac{\exp(-\frac{(\ln(y)-\mu \ln(\Delta))^2}{2\sigma^2\ln^2(\Delta)})}{\sqrt{2\pi}\sigma y \ln(\Delta)}
\end{equation*}

In order to find the ensemble averaged TAMSD, $\langle \hat{\delta^2}\rangle$ simply calculate the average $y$. We find the result of equation \ref{eqn:NDmean}:
\begin{equation*}
\Delta^{\mu}  \cdot \exp(\frac{1}{2}\sigma^2 \ln^2(\Delta))
\end{equation*}

While we assume a normal distribution for the anomalous exponent, one must take care due to the limitation of $\alpha>0$. It can be shown that for a random variable $\alpha$ normally distributed with mean $\mu$ and standard deviation $\sigma$, the limitation of $\alpha>0$ causes the true mean of the distribution to become $\tilde{\mu}=\mu+\frac{(\sqrt{2/\pi}\sigma)\exp(\mu^2/(2\sigma^2))}{1 +Erf[\mu/(\sqrt{2}\sigma)]}$. Where $Erf$ is the error function.

Taking again $y=\Delta^\alpha$, then the average $y$ is $\hat{y}=\Delta^{\mu}\cdot\exp[\sigma_\alpha^2 \ln(\Delta)]\cdot\frac{1+Erf[\frac{\mu+\sigma^2\ln(\Delta)}{\sqrt{2}\sigma}]}{1+Erf[\frac{\mu}{\sqrt{2}\sigma}]}$. We see that we have received the same value as in equation \ref{eqn:NDmean} multiplied by a correction term,$C$. This correction term is greater than unity for all $\Delta>0$. However, unless $\mu\sim\sigma$, it's influence is extremely small. 

In more detail, designate $a\equiv\mu/(\sqrt{2}\sigma)$ and $b\equiv\sigma\ln(\Delta)/\sqrt{2}$. Remembering that $Erf(a+b)= Erf(a)+\frac{2e^{-a^2}b}{\sqrt{\pi}}+O[b^2]$ we find that $C\approx 1+\dfrac{2e^{-a^2}b}{\sqrt{\pi}(1+Erf(a))}$. Therefore, the correction diminishes exponentially with the ratio of $\mu$ to $\sigma$.

\section*{References}
\bibliography{errorsRef}

\section*{Figure Legends}

\end{document}